\documentclass[11pt]{article} 
\usepackage[utf8]{inputenc}

\usepackage{amsfonts}
\usepackage{amsmath}
\usepackage{amssymb}
\usepackage{amsthm}
\usepackage{caption}
\usepackage{color}
    \definecolor{darkgreen}{rgb}{0,0.5,0}
    \definecolor{darkblue}{rgb}{0,0,0.6}
    \definecolor{purple}{rgb}{0.4,.2,0.7}
\usepackage[margin = 2.5cm]{geometry}
\usepackage{graphicx}
\usepackage{array}
\usepackage[hyperfootnotes = false, colorlinks = true, linkcolor = darkblue, citecolor = blue,urlcolor  = darkblue]{hyperref}
\usepackage{subcaption}
\usepackage{appendix}
\usepackage{braket}
\usepackage{tikz}
\usepackage{mathrsfs} 
\allowdisplaybreaks[4]

\begin{document}
\begin{titlepage}
	\renewcommand{\thefootnote}{\fnsymbol{footnote}}
\thispagestyle{empty}
\begin{center}
    ~\vspace{30mm}
    
    {\Large \bf { Recovery of consistency
  in thermodynamics of regular black holes in Einstein's gravity coupled with nonlinear  electrodynamics}}

    \vspace{0.7in}
    
     {\bf Yang Guo${}^{}$\footnote{guoy@mail.nankai.edu.cn}, Hao Xie${}^{}$\footnote{xieh@mail.nankai.edu.cn}, and Yan-Gang Miao${}^{}$\footnote{Corresponding author: miaoyg@nankai.edu.cn}}

   \vspace{0.2in}

   {\em School of Physics, Nankai University, Tianjin 300071, China}
     
    \vspace{0.5in}

\end{center}

\begin{abstract}
 As one of candidate theories in the construction of regular black holes, Einstein's gravity coupled with nonlinear electrodynamics  has been a topic of great concerns. Owing to the coupling between Einstein's gravity and nonlinear electromagnetic fields, we need to reconsider the first law of thermodynamics, which will lead to a new thermodynamic phase space. In such a phase space, the equation of state accurately describes the complete phase transition process of regular black holes. The Maxwell equal area law strictly holds when the phase transition occurs, and the entropy  obeys the Bekenstein-Hawking area formula, which is compatible with the situation in Einstein's gravity.

\end{abstract}

\vspace{1in}


\setcounter{tocdepth}{3}


\end{titlepage}
\tableofcontents
\renewcommand{\thefootnote}{\arabic{footnote}}
\setcounter{footnote}{0}
\section{Introduction} 
Regular black holes in Einstein's gravity coupled with nonlinear  electrodynamics have been studied~\cite{Ayon-Beato:1998hmi,Ayon-Beato:1999kuh,Bronnikov:2000vy,Okyay:2021nnh} extensively in recent years, which is related closely to the singularity of spacetime. In Einstein's gravity, the singularity theorem established by Hawking and Penrose~\cite{hawking1973large} claims that it is an inevitable feature of the theory itself. However, it is generally accepted that such a singularity is unphysical on the classical level and should be served as an evidence that the general relativity requires modifications or generalizations to incorporating quantum theories. The nonlinear  electrodynamics provides a crucial contribution to avoiding singularities of black holes, which allows us to construct~\cite{Ayon-Beato:2000mjt,Burinskii:2002pz,Ayon-Beato:2004ywd,Hassaine:2008pw,Balart:2014cga,Fan:2016hvf} regular black holes with a nonlinear electromagnetic source.

More recently, a remarkable progress has been made~\cite{Cai:2004eh,Myung:2008eb,Gonzalez:2009nn,Wang:2018xdz,Wang:2019kxp,Bokulic:2021dtz,Hendi:2012um,Zou:2013owa} in the study of thermodynamics of regular black holes in Einstein's gravity coupled with nonlinear  electrodynamics. However, there exist some difficulties and challenges, for instance, the Maxwell equal area law  is violated~\cite{Fan:2016rih} during the process in the small-large black hole phase transition, and the entropy does not obey~\cite{Tzikas:2018cvs,Rizwan:2020bhp,NaveenaKumara:2020jmk} the Bekenstein-Hawking area formula. After an in-depth review, we find that these difficulties and challenges are caused by a superficial application of the methodology developed~\cite{Kubiznak:2012wp,Altamirano:2013ane,Kubiznak:2016qmn} in Einstein's gravity to the theory of Einstein's gravity coupled with nonlinear electrodynamics. More precisely, the first law of black hole thermodynamics in Einstein's gravity should be modified in the theory of Einstein's gravity coupled with nonlinear electrodynamics. If not, the naive application would lead to an incorrect and incomplete phase space in which the equation of state could not accurately describe the phase transition of regular black holes. Consequently, the first law of thermodynamics should be reconsidered  for regular black holes in the theory of Einstein's gravity coupled with nonlinear  electrodynamics.

After giving an improved first law of thermodynamics in terms of a covariant approach~\cite{Rasheed:1997ns,Zhang:2016ilt}, we can eliminate the inconsistency previously appeared in thermodynamics of regular black holes with the price of introducing a new thermodynamic phase space. In this new phase space, the Maxwell equal area law holds strictly, and the entropy obeys the Bekenstein-Hawking area formula, which is compatible with the situation in Einstein's gravity. In addition, the phase transition behavior of regular black holes is completely described by the equation of state in the new phase space. We may say that our work provides a refined approach to thermodynamics of regular black holes in Einstein's gravity coupled with nonlinear  electrodynamics.
 
Our paper is organized as follows. In Sec.~\ref{sec:NLD}, we introduce the basics for  Einstein's gravity coupled with nonlinear  electrodynamics. In Sec.~\ref{sec:1stlaw}, we derive the first law  and Smarr formula in terms of a covariant approach for regular black holes in Einstein's gravity coupled with nonlinear  electrodynamics. Next, we demonstrate a general approach to thermodynamics of regular black holes in a new extended phase space in Einstein's gravity coupled with nonlinear  electrodynamics in Sec.~\ref{sec:app}. Then we apply this approach to two specific models, Bardeen and Hayward AdS black holes, and investigate their first laws and the Bekenstein-Hawking area formulas in Sec.~\ref{sec:thermodynamics}. Moreover, we make a detailed discussion to a ratio of critical relations for some other black hole models in Sec.~\ref{sec:number}. Finally, we summarize our results in Sec.~\ref{sec:con}.

\section{Einstein's gravity coupled with nonlinear  electrodynamics}
\subsection{Nonlinear  electrodynamics}
\label{sec:NLD}
The Einstein gravity coupled with nonlinear  electrodynamics  can be described by the action,
 \begin{eqnarray}
	S=\frac{1}{16\pi}\int d^4x\sqrt{- g}\left[{ R}-\mathcal{L}(F) \right],\label{actioneinnonem}
\end{eqnarray}
where $R$ is scalar curvature, $F\equiv\frac{1}{4} F^{ab} F_{ab}$ is electromagnetic invariant,  and $F_{ab}\equiv\partial_aA_b-\partial_bA_a$ is field strength with $A_a$ a vector potential. By varying the action with respect to $A_a$, one obtains the following equation,
\begin{eqnarray}
	\nabla_aG^{ab}=0 \label{eq:Gab},
\end{eqnarray}
where $G^{ab}$ is defined by
\begin{eqnarray}
	G^{ab}\equiv\frac{1}{2}\frac{\partial \mathcal{L}(F)}{\partial F_{ab}}=\frac{1}{4}  \mathcal{L}'(F) F^{ab},
\end{eqnarray} 
and the prime denotes a derivative with respect to $F$.
Then one gives the corresponding electric charge and magnetic charges, respectively,
\begin{eqnarray}
	Q_e&=&\frac{1}{4\pi}\int_{\partial\Sigma}{^*}G_{ab}\\
	Q_m&=&\frac{1}{4\pi}\int_{\partial\Sigma}F_{ab},\label{eq:Qm},
\end{eqnarray}
where ${^*}$  is Hodge star operator given by 
\begin{eqnarray}
	{^*}G_{ab}=\frac{1}{2!} \epsilon _{abcd} G^{cd}. \label{eq:sGab}
\end{eqnarray}
In a stationary spacetime, there exists a timelike Killing vector field, $\xi ^a$, with which the electric and magnetic fields can be expressed by
\begin{eqnarray}
	E_a&=&F_{ab}\xi^b,\\
	H_a&=&-{^*}G_{ab}\xi^b.\label{eq:Ha}
\end{eqnarray}

Considering the fact that the field strength, $F_{ab}$, is closed, one has
\begin{eqnarray}
	0=\xi^c \nabla_{[a} F_{bc]}=\frac{1}{6} \xi^c ( \nabla_a F_{bc} + \nabla_b F_{ca} + \nabla_c F_{ab} -\nabla_a F_{cb} - \nabla_b F_{ac} - \nabla_c F_{ba}). \label{eq:xFab}
\end{eqnarray}
Along the Killing vector field, the Lie derivative reads
\begin{eqnarray}
	\mathscr{L}_{\xi} F_{ab}=\xi ^c \nabla_c F_{ab} +F_{cb} \nabla_a \xi ^c +F_{ac} \nabla_b \xi ^c=0. \label{eq:LieFab}
\end{eqnarray}
Substituting Eq.~(\ref{eq:xFab}) into Eq.~(\ref{eq:LieFab}) and considering the  anti-symmetry of $F_{ab}$, one derives the equation,
\begin{eqnarray}
	\xi^c \nabla_a F_{bc} -\xi^c  \nabla_b F_{ac} +F_{bc} \nabla_a \xi^c  - F_{ac} \nabla_b \xi^c=0,
\end{eqnarray}
which can be rewritten, if one utilizes the Leibniz law, as follows:
\begin{eqnarray}
	\frac{1}{2} \left\{\nabla_a (F_{bc} \xi^c) - \nabla_b (F_{ac} \xi^c)\right\} =\frac{1}{2} (\nabla_a E_b - \nabla_b E_a)=\nabla_{[a} E_{b]}=\frac{1}{2}(\text{d} E)_{ab}=0,
\end{eqnarray}
where d is exterior derivative operator. The above relation shows that the electric field, $E_a$, is closed.

Utilizing the definition of exterior derivatives and Hodge duals, one gives
\begin{eqnarray}
	(\text{d}{^*}G)_{fgd}=\frac{3}{2} \nabla_{[f} \epsilon_{gd]ab} G^{ab}. \label{eq:dsGab}
\end{eqnarray}
Contracting both sides of Eq.~(\ref{eq:dsGab}) with $\epsilon^{efgd}$, one further derives
\begin{eqnarray}
	\epsilon^{efgd} (\text{d}{^*}G)_{fgd}=\frac{3}{2} \epsilon^{efgd}  \nabla_{[f} \epsilon_{gd]ab} G^{ab}=6 \nabla_f G^{fe}. \label{eq:edsGab}
\end{eqnarray}
Again contracting both sides of Eq.~(\ref{eq:edsGab}) with $\epsilon_{ecab}$ and using Eq.~\eqref{eq:Gab}, one finally obtains
\begin{eqnarray}
	(\text{d}{^*}G)_{cab}=- \epsilon_{ecab} \nabla_f G^{fe}=0,
\end{eqnarray}
which shows that  ${^*}G_{ab}$ is closed.
As a result, the electric field, $E_a$, and magnetic field, $H_a$, can be expressed as the exterior derivatives of the scalar fields $\Phi$ and $\Psi$, respectively, 
\begin{eqnarray}
	E_a&=&- \nabla_a \Phi,\\
	H_a&=&-\nabla_a \Psi,\label{eq:Psi]}
\end{eqnarray}
where $\Phi$ is electric potential and $\Psi$ magnetic potential.

	\subsection{First law  and Smarr formula for regular black holes}
	\label{sec:1stlaw}
	We investigate a  model with generality in Einstein's gravity coupled with  nonlinear  electrodynamics, where the Lagrangian density of nonlinear  electrodynamics takes the form~\cite{Fan:2016hvf},
	\begin{eqnarray}
		\mathcal{L}(\mathcal{F})=\frac{4\mu}{\alpha} \frac{ (\alpha\mathcal{F} )^{(\nu+3)/4}}{ \left[ 1+(\alpha\mathcal{F})^{\nu/4}\right] ^{(\mu+\nu)/\nu}}\label{eq:lagFW},
	\end{eqnarray}
$\mu$ and $\nu$ are dimensionless constants which  characterize the  nonlinear degree of  electromagnetic fields, $\mathcal{F}$ is electromagnetic invariant,  and $\alpha$ is constant parameter with the dimension of length squared. In this model, there existed some misinterpretations for mass parameters although such a model gave contributions to the study of regular black holes. In Ref.~\cite{Fan:2016hvf} the ADM mass was defined by
\begin{eqnarray}
	M_{\rm ADM}=M+\frac{q^3}{\alpha},
\end{eqnarray}
where $q$ was magnetic charge, and $M$ was interpreted improperly as the gravitational mass, i.e., $M=M_g$. Such a setup gives rise to an imperfect result that the corresponding regular black hole was non-singular only with a vanishing gravitational mass.
We note that a regular black hole with a vanishing gravitational mass is acceptable mathematically but ill-defined physically due to lack of gravitational interaction.
 A reasonable interpretation for mass parameters suggests~\cite{Toshmatov:2018cks,Bronnikov:2017tnz} that  the gravitational mass $M_g$  should equal the electromagnetically induced mass $M_{em}$,
 \begin{eqnarray}
	M_g=M_{em}=\frac{q^3}{\alpha},
\end{eqnarray}
which allows the mass parameter of regular black holes to be well explained. In order to  make regular black holes well-defined, that is, the mass parameter that appears in the Lagrangian density denotes ADM mass, we make the crucial transformations in Eq.~\eqref{eq:lagFW},
\begin{eqnarray}
\mathcal{F}\rightarrow\frac{2M}{q} F,\qquad \alpha\rightarrow \frac{q^3}{M},\label{crutransFa}
\end{eqnarray}
and then change the Lagrangian density of nonlinear electrodynamics to the following form, 
	\begin{eqnarray}
	\mathcal{L}(F)=\frac{4 \mu  M \left(2 F q^2\right)^{(\nu +3)/4}}{q^3 \left[ 1+\left(2F q^2\right)^{\nu /4}\right]^{(\mu +\nu)/\nu }},\label{eq:L}
\end{eqnarray}
in which $M$ is exactly the ADM mass as we shall see. We can thus construct regular black holes, such as the Bardeen-like and Hayward-like ones, by varying $\mu$ and $\nu$. As a result, according to Eqs.~(\ref{actioneinnonem}) and (\ref{eq:L}), we deduce a general static and spherically symmetric solution in Einstein's gravity coupled with nonlinear electrodynamics, where the corresponding shape function reads
\begin{eqnarray}
	f=1-\frac{2Mr^{\mu-1}}{(r^\nu+q^{\nu})^{\mu/\nu}}.\label{eq:frz}
\end{eqnarray}

One form of the first law was  established~\cite{Rasheed:1997ns} via a covariant approach in the framework of Einstein's gravity coupled with nonlinear electrodynamics. However, the electromagnetic invariant $F$ was assumed~\cite{Rasheed:1997ns} to be the only variable in the Lagrangian density of nonlinear electrodynamics, see Eq.~(\ref{eq:L}), which leads to the limitation that this form of the first law holds for Born-Infeld black holes but not for Bardeen black holes. As is known, a correct first law should be universal, which means that it holds for any regular black hole solution from this framework.  For the purpose to make this covariant approach valid for any regular black hole in Einstein's gravity coupled with nonlinear electrodynamics, it was suggested~\cite{Zhang:2016ilt} that all parameters in the Lagrangian density of nonlinear electrodynamics, including the electromagnetic invariant and other parameters such as charge and mass, must be treated as variables. As a result, the improved first law was written~\cite{Zhang:2016ilt} as follows:
\begin{eqnarray}
	\delta M=\frac{\kappa}{8\pi} \delta A+\Phi \delta Q_e + \Psi \delta Q_m + \sum_{i} K_i\delta\beta_i,\label{impfirlaw}
\end{eqnarray}
where $\beta_i$ denotes the parameters except the electromagnetic invariant, e.g. mass and charge, and  $K_i$ is given by
\begin{eqnarray}
	K_i=\frac{1}{16\pi}\int_{\Sigma}\frac{\partial \mathcal{L}}{\partial\beta_i}\xi^d\epsilon_{dabc}.\label{eq:Ki}
\end{eqnarray}
Moreover, the field strength can be expressed~\cite{Ayon-Beato:2000mjt} as
\begin{eqnarray}
	F_{ab}=q\sin\theta  (\text{d}\theta)_a \wedge (\text{d}\phi)_b\label{eq:Fab},
\end{eqnarray}
and correspondingly the magnetic charge $Q_m$ appeared in Eq.~(\ref{impfirlaw}) can be verified as $q$,
\begin{eqnarray}
	Q_m=\frac{1}{4\pi}\int_{\partial\Sigma}F_{ab}=q,
\end{eqnarray}
when Eqs.~\eqref{eq:Qm} and \eqref{eq:Fab} are utilized.

Now we focus on the improved form of the first law, Eq.~(\ref{impfirlaw}), and its corresponding Smarr formula for the general model depicted by Eqs.~(\ref{actioneinnonem}) and (\ref{eq:L}). At first, it is obvious that  the electric potential, $\Phi$, vanishes because this model is irrelevant to electric charge. Then, we compute the magnetic field by using Eq.~\eqref{eq:Ha},
\begin{eqnarray}
		H_a=\frac{\mu  M   \left[\nu +3-(\mu -3) q^{\nu } r^{-\nu }\right]q^{\nu-1 } r^{1-\nu }}{2 r^2 \left(1+q^{\nu } r^{-\nu }\right)^{\mu/\nu +2}} (\text{d}r)_a,
\end{eqnarray}
and consequently obtain the magnetic potential on the horizon by substituting the above equation into Eq.~\eqref{eq:Psi]},
\begin{eqnarray}
	\Psi=\frac{M  \left\{\mu  q^{\nu } r_+^{-\nu }+3 \left( 1+ q^{\nu } r_+^{-\nu }\right) \left[\left(1+q^{\nu } r_+^{-\nu }\right)^{\mu /\nu }-1\right]\right\}}{2 q \left(1+q^{\nu } r_+^{-\nu }\right)^{(\mu +\nu)/\nu }}.\label{Psi}
\end{eqnarray}
Next, considering that $\beta_i$ corresponds to $q$ and $M$ in the model described by Eqs.~(\ref{actioneinnonem}) and (\ref{eq:L}), we derive the corresponding $K_i$ by using Eq.~\eqref{eq:Ki},
\begin{eqnarray}
	K_q&=&\frac{1}{16\pi}\int_{\Sigma}\frac{\partial \mathcal{L}}{\partial q}\xi^d\epsilon_{dabc}= \frac{M r_+^{\mu } \left[3-3 \left(q^{\nu } r_+^{-\nu }+1\right)^{\mu /\nu }+\mu  q^{\nu }(r_+^{\nu }+ q^{\nu })^{-1}\right]}{2 q \left(r_+^{\nu }+ q^{\nu }\right)^{\mu /\nu }},\label{Kq}\\
	K_M&=&\frac{1}{16\pi}\int_{\Sigma}\frac{\partial \mathcal{L}}{\partial M}\xi^d\epsilon_{dabc}=1-\frac{r_+^{\mu }}{\left(r_+^{\nu }+q^{\nu }\right)^{\mu /\nu }}.\label{KM}
\end{eqnarray}
At this stage, we deduce the first law of mechanics for regular black holes in Einstein's gravity coupled with nonlinear electrodynamics,
\begin{eqnarray}
	\delta M=\frac{\kappa}{8\pi} \delta A+\Psi \delta q +  K_q \delta q +K_M \delta M,\label{1stlaw}
\end{eqnarray}
and the corresponding Smarr formula,
\begin{eqnarray}
M=\frac{\kappa}{4\pi} A+ \Psi q + K_q q + K_M M.
\end{eqnarray}
By varying $\mu$ and $\nu$, we can check that the Bardeen and Hayward black holes satisfy the above first law and Smarr formula. Here we note that the mass parameter $M$ that appears in Eq.~(\ref{eq:L}) is exactly the ADM mass, which makes the mass parameter  well-explained for regular black holes.
		
\section{Thermodynamics in a new extended phase space}
\subsection{Thermodynamics in a new extended phase space for regular black holes in Einstein's gravity coupled with nonlinear electrodynamics}
\label{sec:app}
		\setcounter{equation}{0}
		
In  Einstein's gravity, the thermodynamics of black holes has been studied~\cite{Altamirano:2013uqa,Miao:2017fqg,Wei:2015iwa,Zhang:2015ova} extensively and many effective  methodologies have been developed~\cite{Kubiznak:2012wp,Altamirano:2013ane,Kubiznak:2016qmn}. However, in Einstein's gravity coupled with nonlinear  electrodynamics, the study of thermodynamics of regular black holes  has presented difficulties and problems, such as inconsistencies~\cite{Fan:2016rih} between the Maxwell equal area law and the Gibbs free energy during the phase transition of small-large regular black holes, and deviations~\cite{Tzikas:2018cvs} in calculations of thermodynamic quantities. We point out that one cannot simply generalize the  methodology developed in Einstein's gravity to Einstein's gravity coupled with nonlinear electrodynamics for the study of regular black hole thermodynamics. In particular, we must reconsider thermodynamics based on an improved first law we are going to figure out. In this subsection, we provide the fundamental approach to investigate  thermodynamics of regular black holes in Einstein's gravity coupled with nonlinear electrodynamics.  In a new extended phase  space where the cosmological constant is identified as pressure,  $P=-\Lambda/(8\pi)$, we write an improved first law of thermodynamics from Eq.~\eqref{1stlaw} as a compact form,
\begin{eqnarray}
d M=T' d S+\Psi' d q + V' d P, \label{1stLaw}
\end{eqnarray}
where 
\begin{eqnarray}
T' = \frac{T}{1-K_M},\qquad \Psi' = \frac{\Psi+K_q}{1-K_M},\qquad V'=\frac{V}{1-K_M}.\label{tpvpto}
\end{eqnarray}
We note that the new extended phase space consists of ($T'$,  $S$), ($\Psi'$, $q$), and ($P$, $V'$) rather than ($T$,  $S$), ($\Psi$, $q$), and ($P$, $V$). In other words, it is composed  of ($T'$, $\Psi'$, $P$, $S$, $q$, $V'$).  As we shall see in the next subsection, it is crucial to fix ($T'$, $\Psi'$, $V'$) for addressing  those  difficulties mentioned above in thermodynamics of regular black holes in Einstein's gravity coupled with nonlinear electrodynamics. 
		
In thermodynamics, the Gibbs free energy is equal to the maximum work that a thermodynamically closed system performs at constant temperature and pressure. Therefore, we can define the new Gibbs free energy by using the new temperature $T'$ instead of the Hawking temperature $T$,
\begin{eqnarray}
G' = M-T' S, \label{eq:G}
\end{eqnarray}
in order to distinguish from $G = M-T S$. We emphasize that the newly defined Gibbs free energy is distinct from the off-shell generalized free energy defined~\cite{Eune:2013qs,Li:2020khm,Li:2022oup,Yang:2021ljn} by an ensemble temperature. The reason is that an ensemble temperature is an external adjustable parameter and it is independent of black hole parameters. Here $T'$ is correlated highly with the Hawking temperature and depends on black hole parameters. More importantly, it satisfies the equation of state. Thus the new Gibbs free energy defined in Eq.~\eqref{eq:G} is on-shell. Using Eqs.~(\ref{1stLaw}) and (\ref{eq:G}), we obtain the differential form of  Gibbs free energy,
\begin{eqnarray}
d G' = -S d T'+ \Psi' d q + V' d P.  \label{eq:dG}
\end{eqnarray}

\begin{figure}[h]
	\begin{subfigure}{.5\textwidth}
		\centering
		\includegraphics[width=.9\linewidth]{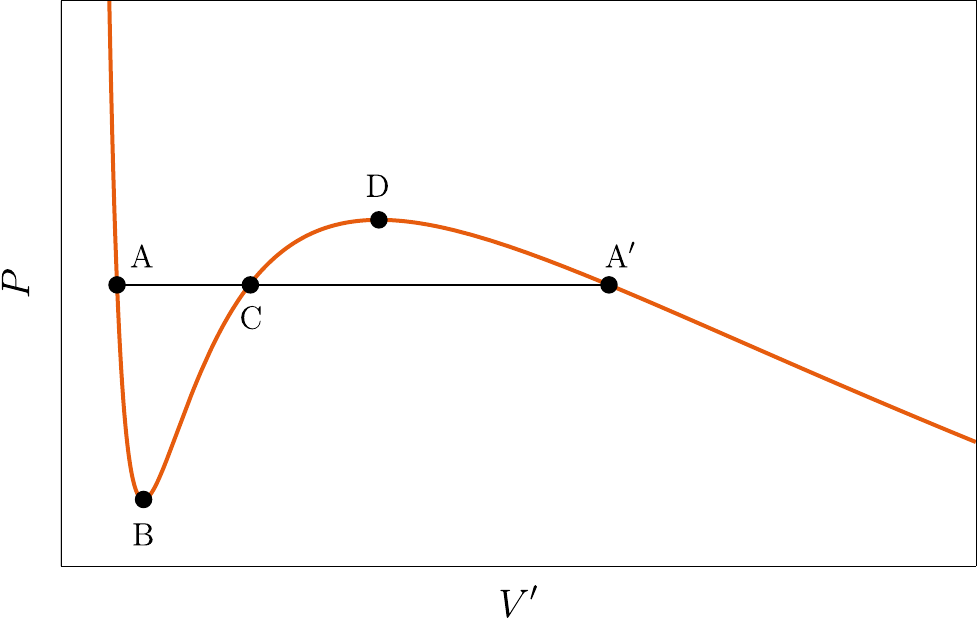}  
		\caption{$P-V'$ plane}\label{sbfig:PV}
	\end{subfigure}
	\begin{subfigure}{.5\textwidth}
		\centering
		\includegraphics[width=.65\linewidth]{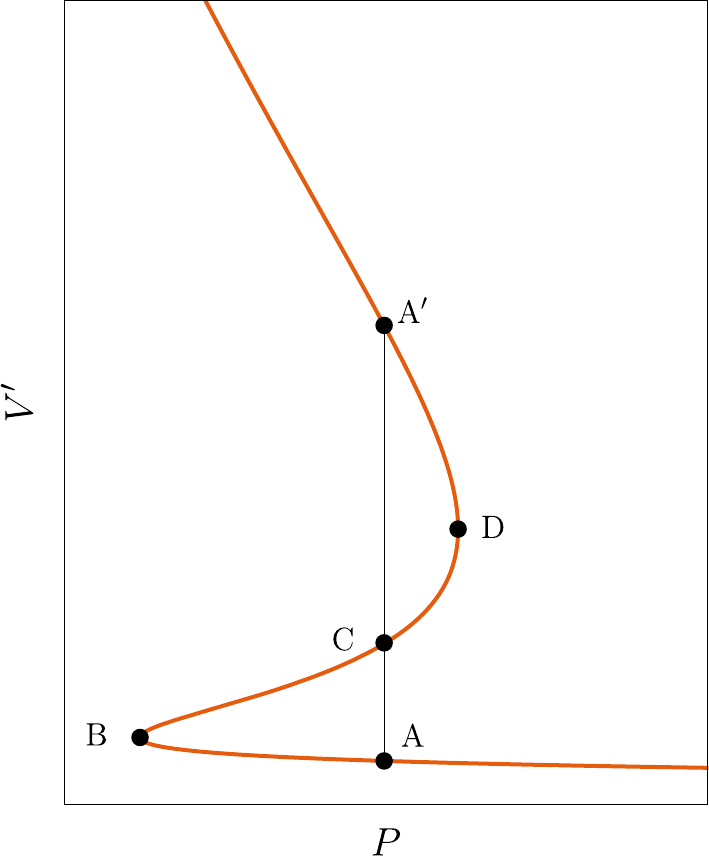}  
		\caption{$V'-P$ plane}
	\end{subfigure}
	\caption{The oscillatory behavior of regular black holes in Einstein's gravity coupled with nonlinear electrodynamics.}
	\label{fig:PV}
\end{figure}

\begin{figure}[h]
	\begin{subfigure}{.5\textwidth}
		\centering
		\includegraphics[width=.9\linewidth]{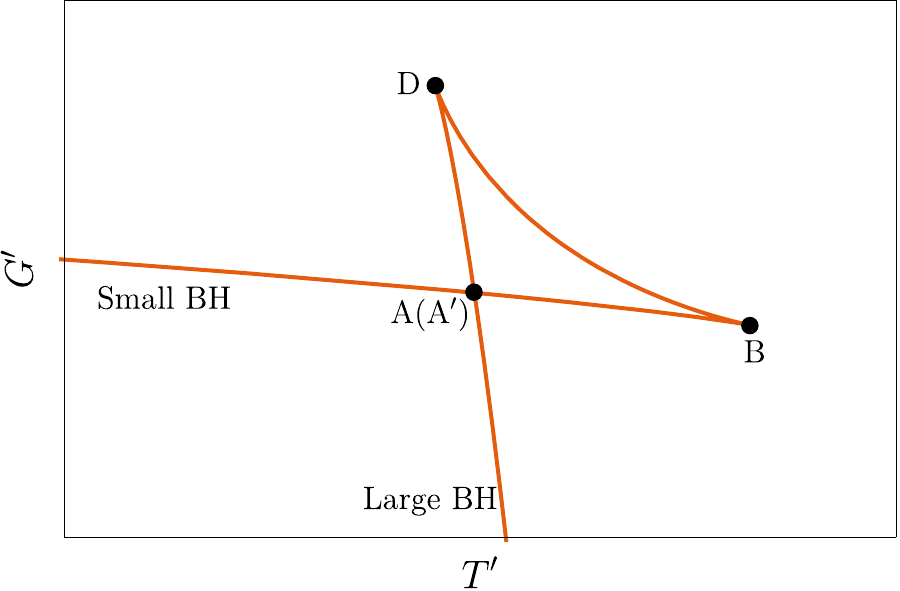}  
		\caption{$G'-T'$ plane}\label{sbfig:GT}
	\end{subfigure}
	\begin{subfigure}{.5\textwidth}
		\centering
		\includegraphics[width=.9\linewidth]{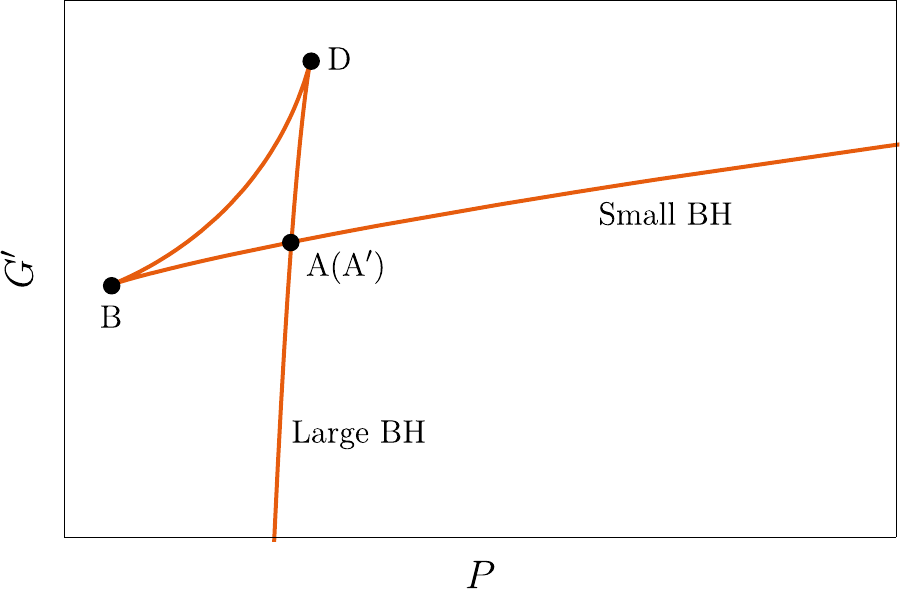}  
		\caption{$G'-P$ plane}
	\end{subfigure}
	\caption{The Gibbs free energy $G'$ with respect to $T'$ and $P$ for regular black holes in Einstein's gravity coupled with nonlinear electrodynamics.}
	\label{fig:GTGP}
\end{figure}

In order to clarify the relation between phase transitions and the Maxwell equal area law, we plot the oscillatory behavior on the plane $(P, V')$ in Fig.~\ref{fig:PV}, and the relations of $(G', T')$ and $(G', P)$ in Fig.~\ref{fig:GTGP}. When a regular black hole transforms from state A to state ${\rm A'}$, the two states have the same Gibbs free energy.  Therefore,  by integrating Eq.~\eqref{eq:dG} at constant $T'$ and $q$, we have
\begin{eqnarray}
\oint V'dP=0,\label{eq:equarea}
\end{eqnarray} 
where the integral is over the closed path:   A $\rightarrow$ B $\rightarrow$ C $\rightarrow$ D $\rightarrow$ ${\rm A'}$ in Fig.~\ref{fig:PV}, where the pressure keeps unchanged from state A to state ${\rm A'}$. Thus, we have the equal area law,
\begin{eqnarray}
\rm Area(ABC) = Area(A'DC).
\end{eqnarray}

\begin{figure}[htbp]
\begin{subfigure}{.5\textwidth}
\centering
\includegraphics[width=.9\linewidth]{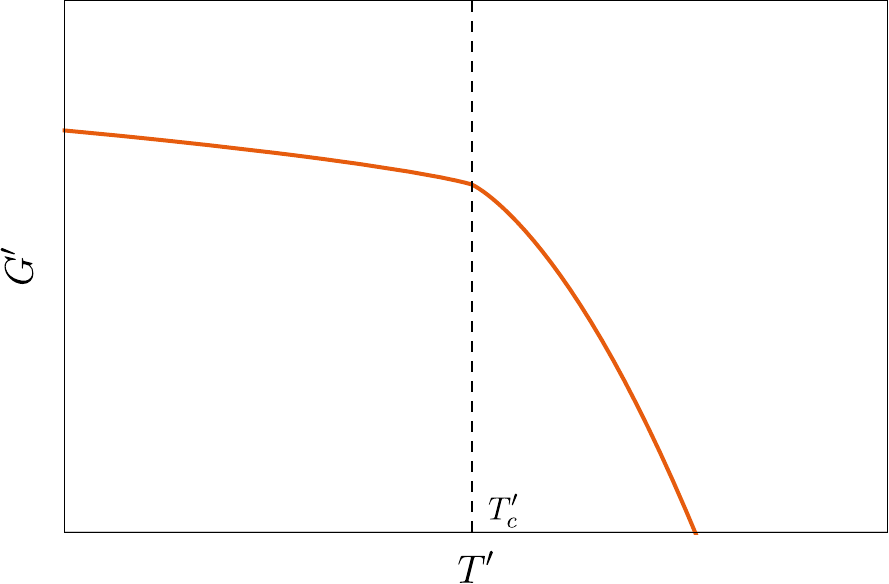}  
\caption{}\label{sbfig:cGT}
\end{subfigure}
\begin{subfigure}{.5\textwidth}
\centering
\includegraphics[width=.9\linewidth]{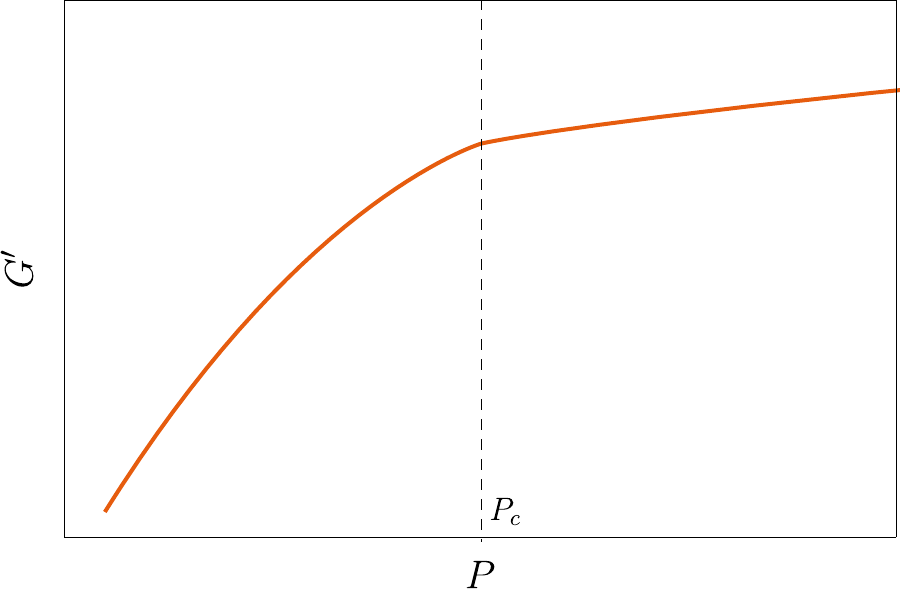}  
\caption{}\label{sbfig:cGP}
\end{subfigure}
\begin{subfigure}{.5\textwidth}
\centering
\includegraphics[width=.9\linewidth]{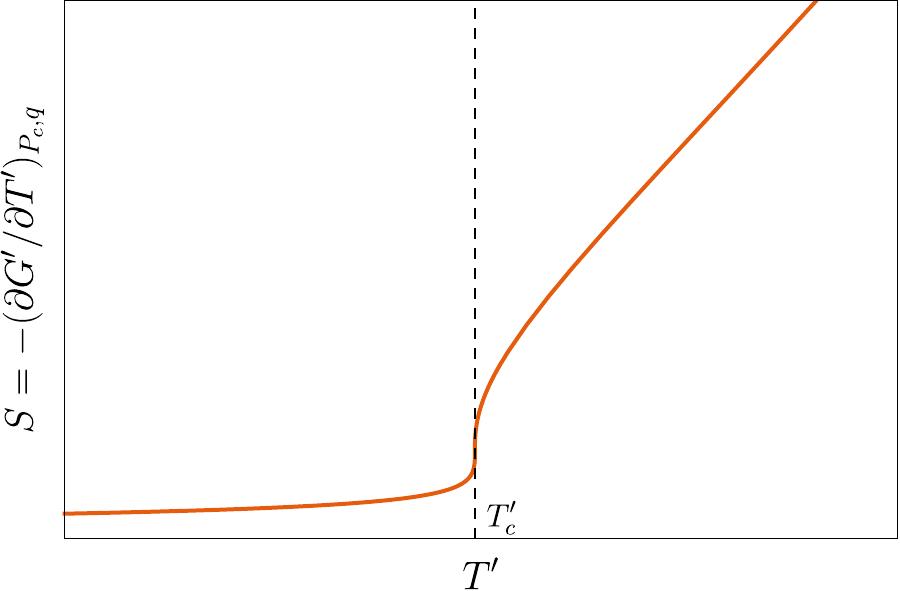}  
\caption{}\label{sbfig:ST}
\end{subfigure}
\begin{subfigure}{.5\textwidth}
\centering
\includegraphics[width=.9\linewidth]{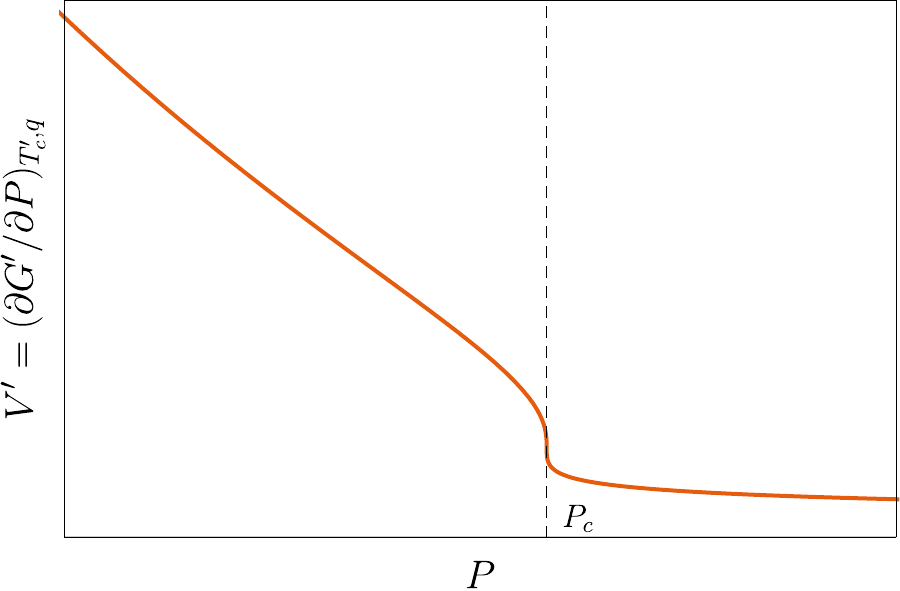}  
\caption{}\label{sbfig:VP}
\end{subfigure}
\begin{subfigure}{.5\textwidth}
\centering
\includegraphics[width=.9\linewidth]{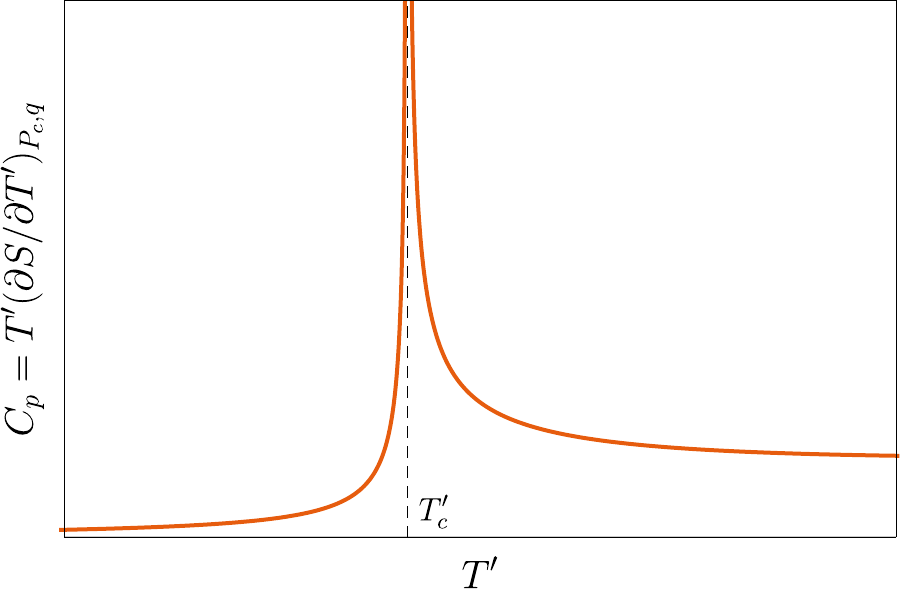}  
\caption{}\label{sbfig:cCpT}
\end{subfigure}
\begin{subfigure}{.5\textwidth}
\centering
\includegraphics[width=.9\linewidth]{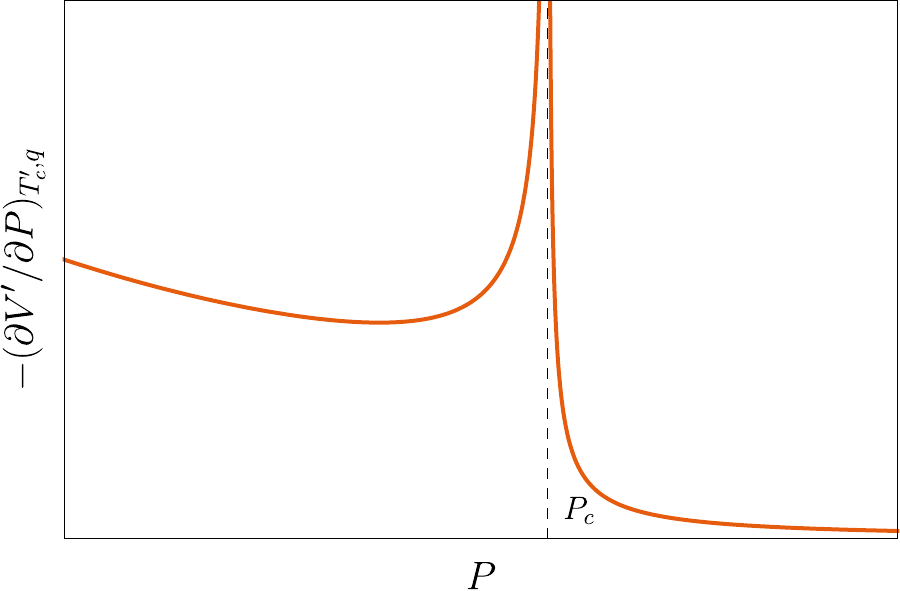}  
\caption{}\label{sbfig:cVPP}
\end{subfigure}
\caption{The thermodynamic behaviors  near the critical points of regular black holes in Einstein's gravity coupled with nonlinear electrodynamics.}
\label{fig:criti}
\end{figure}

The presence of oscillatory behaviors on the $P-V'$ plane characterizes the phase transitions occurring in regular black holes, and so does the presence of swallow tail behaviors on the $G'-T'$ and $G'-P$ planes.
The oscillatory behavior in Fig.~\ref{fig:PV} appears only at the  temperature below the critical value $T'_c$, which shows that the Maxwell equal area law holds.  When $T'$ increases gradually, the area(ABC) or ${\rm area(A'DC)}$ will become small. Meanwhile, the area of swallow tails in Fig.~\ref{fig:GTGP} also becomes small. Once the  temperature or pressure reaches the critical value, $T'_c$ or $P_c$, the oscillatory behavior in Fig.~\ref{fig:PV} and the swallow tail behavior in Fig.~\ref{fig:GTGP} disappear. We plot the critical behavior of $G'$ in Figs.~\ref{sbfig:cGT} and \ref{sbfig:cGP}. At this critical point, the first-order phase transition turns into a second-order one because the entropy and volume are finite and continuous. We write the entropy and volume,
\begin{eqnarray}
S&=&-	\left( \frac{\partial G'}{\partial T'}\right)_{P_c, \,q},\\ \label{eq:S}
V'&=&	\left( \frac{\partial G'}{\partial P}\right)_{T_c', \,q}, \label{eq:V}
\end{eqnarray}
and see their finiteness and continuity from the two sides of $T_c'$ and of $P_c$ in Figs.~\ref{sbfig:ST} and \ref{sbfig:VP}, respectively.
At this second-order phase transition point, the two phases stay at the same thermodynamic state, meaning no coexistence of small and large regular black holes. Nonetheless, the second-order derivatives of $G'$ with respect to $P$ and $T'$ are infinite at this point, 
\begin{eqnarray}
\left( \frac{\partial ^2G'}{\partial P ^2}\right)_{T'_c, \,q} &=& \left( \frac{\partial V'}{\partial P}\right)_{T'_c, \,q} \rightarrow - \infty,\\
 \left( \frac{\partial ^2G'}{\partial T'^2}\right)_{P_c, \,q} &=& -\left( \frac{\partial S}{\partial T'}\right)_{P_c, \,q} \rightarrow - \infty \label{eq:Cp}.
\end{eqnarray}
Therefore, the heat capacity of regular black holes, 
\begin{eqnarray}
C_p=	T' \left( \frac{\partial S}{\partial T'}\right)_{P_c, \,q} =-T' \left( \frac{\partial ^2G'}{\partial T'^2}\right)_{P_c, \,q}\rightarrow + \infty,
\end{eqnarray}
exhibits an infinite peak at $T_c'$ or $P_c$ because it corresponds to the second-order derivative of Gibbs free energy, which is shown in Figs.~\ref{sbfig:cCpT} and \ref{sbfig:cVPP}.

\subsection{Applications to specific regular black hole models}
\label{sec:thermodynamics}

\subsubsection{Bardeen AdS black holes}
\label{sec:Bard}

 Now we revisit the thermodynamics of Bardeen AdS black holes, give the correct thermodynamic quantities and investigate the  phase transition  in an extended phase space. The  action  of Einstein's gravity coupled  with nonlinear electrodynamics in the four-dimensional spacetime reads
 \begin{eqnarray}
 	S=\frac{1}{16\pi}\int d^4x\sqrt{-g}\left( R-2\Lambda-\mathcal{L}(F) \right),
 \end{eqnarray}
  where $\Lambda$ is related to the AdS radius $l$ via the relation, $\Lambda=-3/l^2$,  and $\mathcal{L}(F)$ can be found by evaluating Eq.~\eqref{eq:L} at $\mu=3$ and $\nu=2$. Then the shape function takes~\cite{Ayon-Beato:2000mjt} the form in the metric of Bardeen black holes in the AdS spacetime,
  \begin{eqnarray}
  	f(r)=1-\frac{2Mr^2}{(q^2+r^2)^{3/2}}+\frac{r^2}{l^2}.
  \end{eqnarray}

A number of works have been  devoted~\cite{Tzikas:2018cvs,Rizwan:2020bhp,Pu:2019bxf} to explore thermodynamic properties of Bardeen AdS black holes. However, the main issue is the violation of the Bekenstein-Hawking area formula, where the formula, $S_{\rm BH}=\rm Area/4$, denotes that the entropy of black holes  is proportional to the area of an event horizon in Einstein's gravity. Let us give a brief review on how the issue arises. The reason is that the starting point is an incorrect first law,
 \begin{eqnarray}
 	dM=TdS+ \varphi dq+VdP,
 \end{eqnarray}
 with which the corresponding thermodynamic quantities take~\cite{Tzikas:2018cvs,Rizwan:2020bhp} the forms,
 \begin{eqnarray}
 	T&=&\frac{3r_+^4+l^2(r_+^2-2 q^2)}{4 \pi  r_+ l^2 \left(q^2+r_+^2\right)},\\
 	V&=&\frac{4}{3} \pi  \left(q^2+r_+^2\right)^{3/2},\label{volvari}\\ \varphi&=&\frac{3q(l^2+r_+^2)\sqrt{q^2+r_+^2}}{2l^2r_+^2},\\
	S&=&2 \pi  \left(\frac{r_+}{2}-\frac{q^2}{r_+}\right) \sqrt{q^2+r_+^2}+\frac{3}{2}\pi q^2 \log \left(\sqrt{q^2+r_+^2}+r_+\right)\label{eq:incS}.
\end{eqnarray}
where $r_+$  stands for the horizon radius which is the solution of the algebraic equation, $f(r_+) = 0$. It is clear that Eq.~\eqref{volvari} deviates from the volume of a sphere, and in particular,  Eq.~\eqref{eq:incS} does not obey the Bekenstein-Hawking area formula, which shows that the thermodynamic quantities, ($T, \varphi, P, S,  q,  V$), do not construct a correct and complete phase space. As is known,
Einstein's gravity coupled with nonlinear  electrodynamics is still within the framework of general relativity, and thus the entropy of regular black holes should obey the Bekenstein-Hawking area formula. Let us recover the consistency in
thermodynamics of regular black holes in Einstein's gravity coupled with nonlinear  electrodynamics.

We start with the improved first law given in Sec.~\ref{sec:app}, see Eqs.~(\ref{1stLaw}) and (\ref{tpvpto}), and rewrite it in the following form,
 \begin{eqnarray}
 	d M=T d S+\Psi d q + VdP+  K_q d q +K_M d M,\label{Bard1Law}
 \end{eqnarray}
where  $\Psi$, $K_q$ and $K_M$ take the forms,\footnote{They are calculated in Ref.~\cite{Zhang:2016ilt} for the first time.}
\begin{eqnarray}
	\Psi&=&\frac{3M}{2q}\left[ 1-\frac{r_+^5}{\left(q^2+r_+^2\right){}^{5/2}}\right],\nonumber\\
	 K_q&=& \frac{3M}{2q} \left[\frac{2 q^2r_+^3+r_+^5}{\left(q^2+r_+^2\right){}^{5/2}}-1\right],\nonumber\\ K_M&=&1-\frac{r_+^3}{\left(q^2+r_+^2\right)^{3/2}},\label{psikqkm}
\end{eqnarray}
for Bardeen AdS black holes, i.e., $\mu=3$ and $\nu=2$ are substituted in Eqs.~(\ref{Psi}), (\ref{Kq}), and (\ref{KM}).
Now we can reproduce the Bekenstein–Hawking entropy in terms of the improved first law,
\begin{eqnarray}
	S=\int \frac{1-K_M}{T}dM=\pi r_+^2,
\end{eqnarray}
and simultaneously eliminate the deviation of a sphere volume in Eq.~\eqref{volvari},
\begin{eqnarray}
	V=(1-K_M)\left( \frac{\partial M}{\partial P}\right)_{S,\,q}=\frac{4}{3}\pi r_+^3. 
\end{eqnarray}

Next we give the correct and complete phase space and analyze thermodynamic behaviors for Bardeen AdS black holes. We start directly with  Eqs.~(\ref{1stLaw}) and (\ref{tpvpto}),
and compute $T', V'$, and  $\Psi'$ with the help of Eq.~(\ref{psikqkm}),
\begin{eqnarray}
	T' &=& \frac{T}{1-K_M}=\frac{\sqrt{r_+^2+q^2} \left(8 \pi  P r_+^4+r_+^2-2 q^2\right)}{4 \pi  r_+^4},\nonumber\\
	 V'&=&\frac{V}{1-K_M}=\frac{4}{3} \pi  \left(r_+^2+q^2\right)^{3/2},\nonumber\\
 \Psi' &=& \frac{\Psi+K_q}{1-K_M}=\frac{3 M q}{r_+^2+q^2},\label{eq:BarT}
\end{eqnarray}
and note that the phase space composed of ($T'$, $\Psi'$, $P$, $S$, $q$, $V'$) depends on the nonlinear electrodynamic coupling $q$ and it is just the new extended phase space we expect. In the following we analyze thermodynamic behaviors for Bardeen AdS black holes in terms of this phase space. 

\begin{figure}[ht]
	\begin{subfigure}{.5\textwidth}
		\centering
		\includegraphics[width=.93\linewidth]{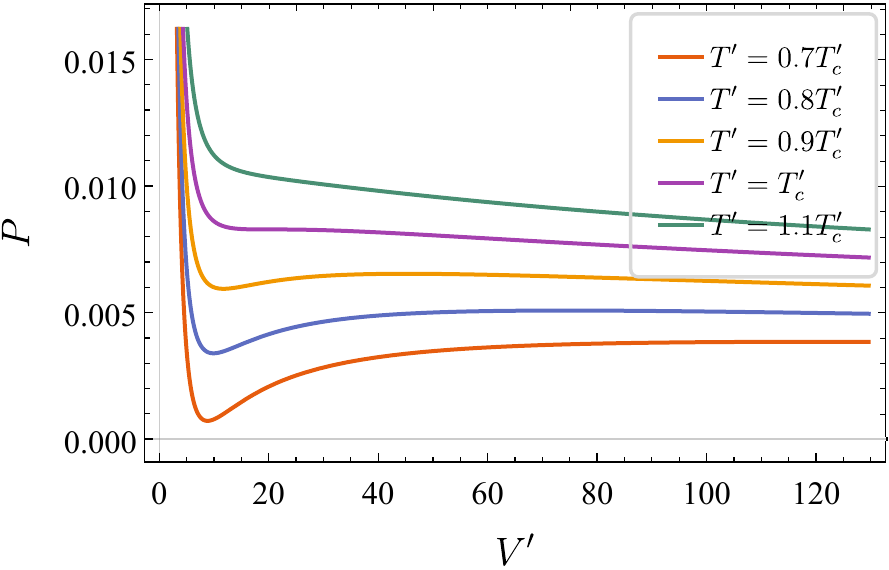}  
		\caption{}
		\label{fig:BarPV}
	\end{subfigure}
\begin{subfigure}{.5\textwidth}
	\centering
	\includegraphics[width=.9\linewidth]{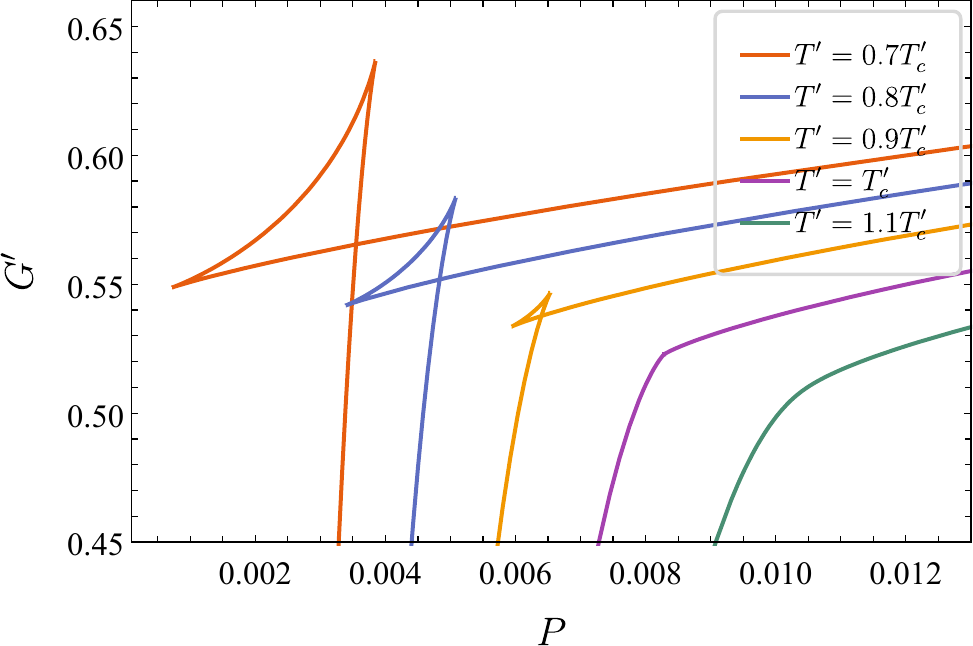}  
	\caption{}
	\label{fig:BarGP}
\end{subfigure}
\caption{(a) The oscillatory behavior on the $P - V'$ plane and (b) the swallow tail behavior on the $G'- P$ plane of Bardeen AdS black holes.}
\label{fig:BarPVGP}
\end{figure}

We express the equation of state as the function of $T'$ and $V'$ instead of $T$ and $V$ in terms of the first two relations of Eq.~(\ref{eq:BarT}),
\begin{eqnarray}
P=	P(T', V')=\frac{12 q^2-\left(\frac{6V'}{\pi }\right)^{2/3}}{2 \pi  \left[\left(\frac{6V'}{\pi }\right)^{2/3}-4 q^2\right]^2}+T'\left(\frac{\pi}{6V' }\right)^{1/3},\label{eq:BarEOS}
\end{eqnarray}
which indeed describes the behaviors of Bardeen AdS black holes as we shall see. Substituting Eq.~\eqref{eq:BarEOS}   into the critical condition with the help of Eq.~\eqref{eq:BarT},
 \begin{eqnarray}
	\left( \frac{\partial P}{\partial V'}\right)_{T'}=0, \qquad \left( \frac{\partial^2 P}{\partial V'^2}\right)_{T'}=0,\label{eq:critical}
\end{eqnarray}
 we obtain the critical values of $T', V'$ and $P$,
\begin{eqnarray}
	T'_c&=&\frac{5 \sqrt{188 \sqrt{10}-505}}{432 \pi  q},\nonumber\\
	V'_c&=&	\frac{4}{3} \left(2 \sqrt{10}+5\right)^{3/2} \pi  q^3,\nonumber\\
	P_c&=&\frac{5 \sqrt{10}-13}{432 \pi  q^2}.
\end{eqnarray}
We emphasize that this critical point coincides strictly with the inflection point at which the first-order phase transition vanishes. As shown in Fig.~\ref{fig:BarPVGP},  $P(T', V')$  describes  the whole process  from a first-order phase transition to a second-order one of Bardeen AdS black holes. If $T'<T'_c$, there exists a small-large black hole phase transition that characterizes the oscillatory behavior of $(P, V')$ in Fig.~\ref{fig:BarPV} and the swallow tail behavior of $(G', P)$ in Fig.~\ref{fig:BarGP}.  If $T'=T'_c$, both the oscillatory behavior and the swallow tail behavior vanish. In particular, when $T'$ and $P$ approach their critical values,  $T'=T'_c$ and $P=P_c$, the small-large black hole phase transition changes into a second-order one. Therefore, we can say that this critical point, $(T'_c, P_c)$, is actually a true second-order phase transition point of Bardeen AdS black holes.

\begin{figure}[ht]
	\begin{subfigure}{.5\textwidth}
		\centering
		\includegraphics[width=.9\linewidth]{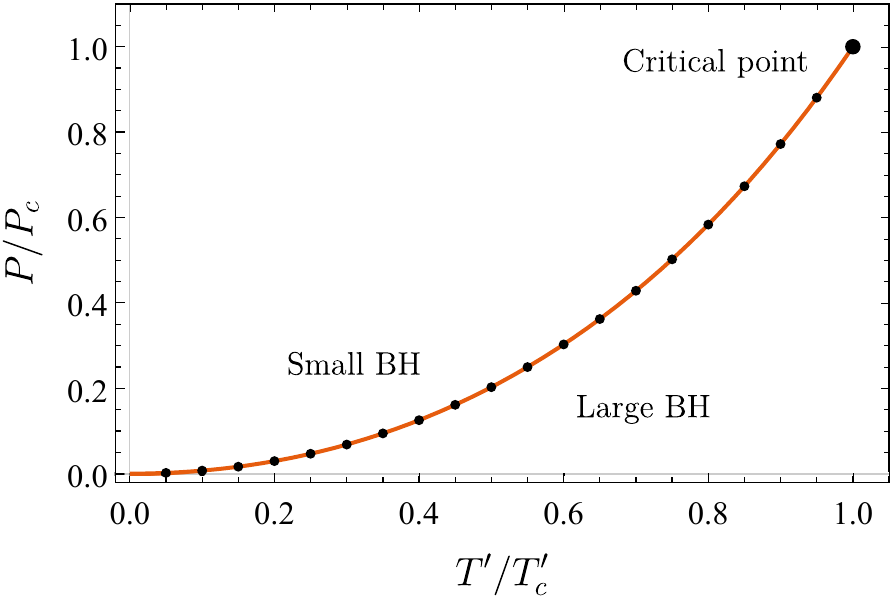}  
		\caption{}
		\label{fig:Bcoe}
	\end{subfigure}
	\begin{subfigure}{.5\textwidth}
		\centering
		\includegraphics[width=.9\linewidth]{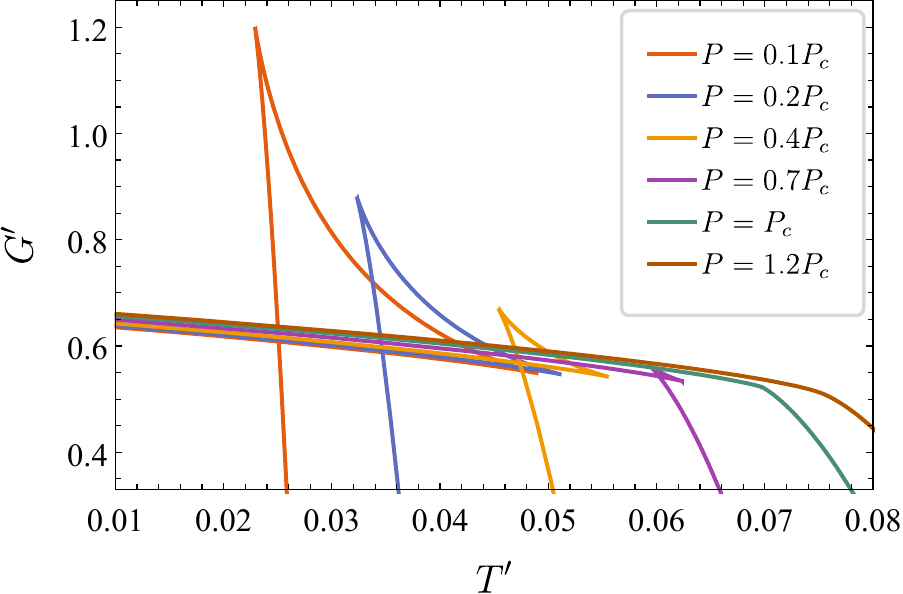}  
		\caption{}
		\label{fig:BarGT}
	\end{subfigure}
	\caption{(a) The coexistence curve of small-large phase and (b) the swallow tail behavior on the $G' - T'$ plane of Bardeen AdS black holes.}
	\label{}
\end{figure}

 In addition, we calculate $P$ and $T'$ of small and large phases of Bardeen AdS black holes, and find that the results given by the new Gibbs free energy Eq.~\eqref{eq:G} are strictly equal to those given by the Maxwell equal area law Eq.~\eqref{eq:equarea}. In Fig.~\ref{fig:Bcoe}, the black dots are computed by the new Gibbs free energy, and the solid curve is given by the Maxwell equal area law. Clearly, the two methods give the same coexistence curve.

The equation of state of  van der Waals fluids reads~\cite{Kubiznak:2012wp}
  	\begin{eqnarray}
  		\left( P+\frac{a}{v^2}\right)(v-b)=kT ,
  	\end{eqnarray}
  	where $k$ is the Boltzmann constant, $v\equiv V/N$ is the specific volume, $N$ is the number of molecules,
  	and the parameter $a$ measures the attraction in the molecules of size $b$.  By the critical condition,
  	\begin{eqnarray}
  		\left( \frac{\partial P}{\partial v}\right)_{T}=0, \qquad \left( \frac{\partial^2 P}{\partial v^2}\right)_{T}=0,\label{Critical}
  	\end{eqnarray}
 we can find the critical point at
  \begin{eqnarray}
  	kT_{\rm c}=\frac{8a}{27b}, \qquad v_{\rm c}=3b, \qquad  P_{\rm c}=	\frac{a}{27b^2}.
  \end{eqnarray}
   It is worth noting the ratio, 
   \begin{eqnarray}
   	\frac{P_cv_c}{kT_c}=\frac{3}{8}.
   	\end{eqnarray}
  In order to compare the Bardeen AdS black hole with the van der Waals fluid, we define the specific volume, $v\equiv 2\sqrt{r_+^2+q^2}$, and rewrite the equation of state as follows:
 \begin{eqnarray}
 	P=	\frac{T'}{v}-\frac{v^2-12 q^2}{2 \pi  \left(v^2-4 q^2\right)^2}.
 \end{eqnarray}
 Then using the critical condition Eq.~\eqref{Critical}, we calculate the critical values,
 \begin{eqnarray}
 	T'_c=\frac{5 \sqrt{188 \sqrt{10}-505}}{432 \pi  q},\qquad v_c=	2 \sqrt{5+ 2 \sqrt{10}}  q,\qquad P_c=\frac{5 \sqrt{10}-13}{432 \pi  q^2},\label{critvaltvp}
 \end{eqnarray}
 and obtain the following ratio at this critical point for the Bardeen AdS black hole, 
 \begin{eqnarray}
 	\frac{P_cv_c}{T'_c}=\frac{2}{5},\label{gennum}
 \end{eqnarray}
 which is clearly independent of $q$ --- the nonlinear electromagnetic coupling. We notice that this ratio is different from that of van der Waals fluids.

\subsubsection{Hayward AdS black holes}
\label{sec:Hay}

The Lagrangian density of Hayward AdS black holes  can be found by evaluating Eq.~\eqref{eq:L}  at $\mu=\nu=3$.
By following the same procedure as that in  Sec.~\ref{sec:Bard}, we give~\cite{Hayward:2005gi} the shape function in the metric of Hayward black holes in the AdS spacetime, 
\begin{eqnarray}
		f=1-\frac{2Mr^{2}}{r^3+q^{3}}+\frac{r^2}{l^2}.
\end{eqnarray}
In Hayward AdS black holes,  the Maxwell equal area law was misinterpreted~\cite{Fan:2016rih}, which gave rise to the law's violation and inconsistency with the Gibbs free energy. The reason is the same as that in Bardeen AdS black holes --- an incorrect first law of thermodynamics was used.

Following the procedure adopted in Sec.~\ref{sec:Bard}, we start with Eq.~(\ref{Bard1Law}) and compute
 $\Psi, K_q$ and $K_M$,
\begin{eqnarray}
		\Psi&=&\frac{3 M q^2 \left(2r_+^3+q^3 \right)}{2 \left(r_+^3+q^3\right)^2}, \nonumber\\
		K_q&=&-\frac{3M q^5}{2\left(r_+^3+q^3\right)^3}, \nonumber\\
		K_M&=&\frac{q^3}{r_+^3+q^3},\label{HpkqkM}
\end{eqnarray}
which correspond to Hayward AdS black holes with $\mu=3$ and $\nu=3$ in Eqs.~(\ref{Psi}), (\ref{Kq}), and (\ref{KM}). Thus we
obtain the Bekenstein–Hawking entropy, 
\begin{equation}
	S=\int \frac{1-K_M}{T}dM=\pi r_+^2,
\end{equation}
and the volume without deviation from a sphere,
\begin{equation}
	V=(1-K_M) \left(\frac{\partial M}{\partial P}\right)_{S, q}=\frac{4 \pi }{3}r_+^3.
\end{equation}

Moreover, starting with 
Eqs.~(\ref{1stLaw}) and (\ref{tpvpto}) together with 
\begin{eqnarray}
T' &=& \frac{8 \pi  P r_+^5+r_+^3-2 q^3}{4 \pi  r_+^4},\nonumber\\
V'&=&\frac{4}{3} \pi  \left(r_+^3+q^3\right),\nonumber\\
\Psi'& =& \frac{3Mq^2r_+^3}{(r_+^3+q^3)^2},\label{tvpsiHayward}
\end{eqnarray}
which can be calculated with the help of Eq.~(\ref{HpkqkM}),
we provide the new extended phase space, ($T'$, $\Psi'$, $P$, $S$, $q$, $V'$), which is similar to that in Bardeen AdS black holes.
Therefore, we express the equation of state by using the first two relations of Eq.~(\ref{tvpsiHayward}) for Hayward AdS black holes as follows:
\begin{equation}
	P=P(T',V')=\frac{4 \pi  T' \left(\frac{3 V'}{4 \pi }-q^3\right)^{4/3}+3 q^3-\frac{3 V'}{4 \pi }}{8 \pi  \left(\frac{3 V'}{4 \pi }-q^3\right)^{5/3}},
\end{equation}
and compute the critical values from Eq.~\eqref{eq:critical}, 
\begin{eqnarray}
		T'_c&=&\frac{3}{8\sqrt[3]{20} \pi  q},\nonumber \\
		V'_c&=&28 \pi  q^3, \nonumber\\
		P_c&=&\frac{3}{80 \sqrt[3]{50}  \pi  q^2}.
\end{eqnarray}

\begin{figure}[ht]
	\begin{subfigure}{.5\textwidth}
		\centering
		\includegraphics[width=.93\linewidth]{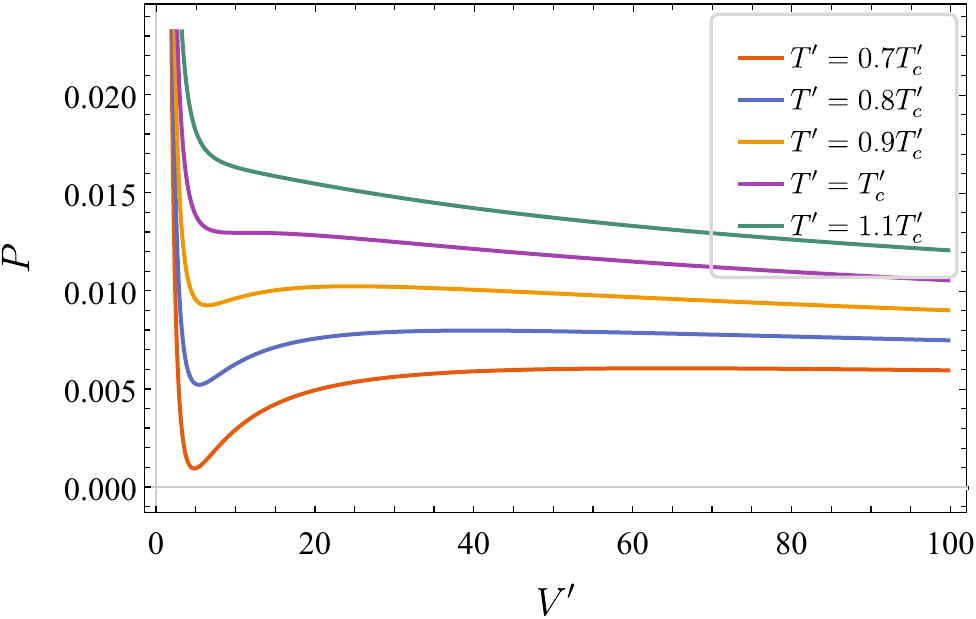}  
		\caption{}
		\label{fig:HayPV}
	\end{subfigure}
	\begin{subfigure}{.5\textwidth}
		\centering
		\includegraphics[width=.9\linewidth]{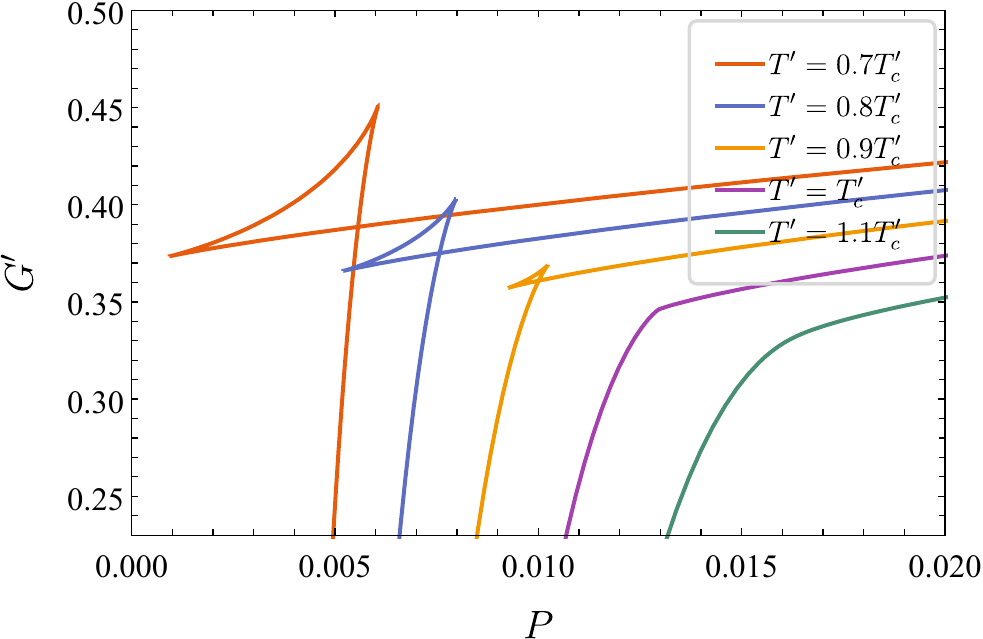}  
		\caption{}
		\label{fig:HayGP}
	\end{subfigure}
	\caption{(a) The oscillatory behavior on the  $P -V'$ plane and (b) the swallow tail behavior on the $G' - P$ plane of Hayward AdS black holes.}
	\label{fig:HayPVGP}
\end{figure}

The critical point coincides with the point at which the first-order phase transition vanishes.  The oscillatory behavior on the $P-V'$ planes is shown in Fig.~\ref{fig:HayPV}. Using our newly defined Gibbs free energy Eq.~\eqref{eq:G}, we plot the swallow tail in Fig.~\ref{fig:HayGP}. When $T'<T'_c$, there exists the small-large black hole phase transition which is characterized by the oscillatory behavior in Fig.~\ref{fig:HayPV} and by the swallow tail in Fig.~\ref{fig:HayGP}. When $T'=T'_c$ and $P=P_c$, the first-order phase transition changes into a second-order one. 

\begin{figure}[ht]
	\begin{subfigure}{.5\textwidth}
		\centering
		\includegraphics[width=.9\linewidth]{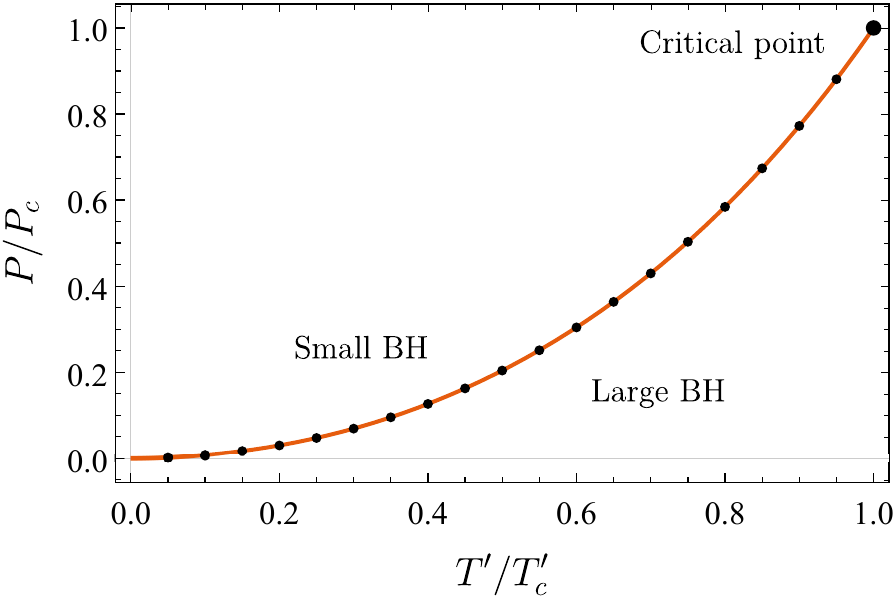}  
		\caption{}
		\label{fig:Hcoe}
	\end{subfigure}
	\begin{subfigure}{.5\textwidth}
		\centering
		\includegraphics[width=.9\linewidth]{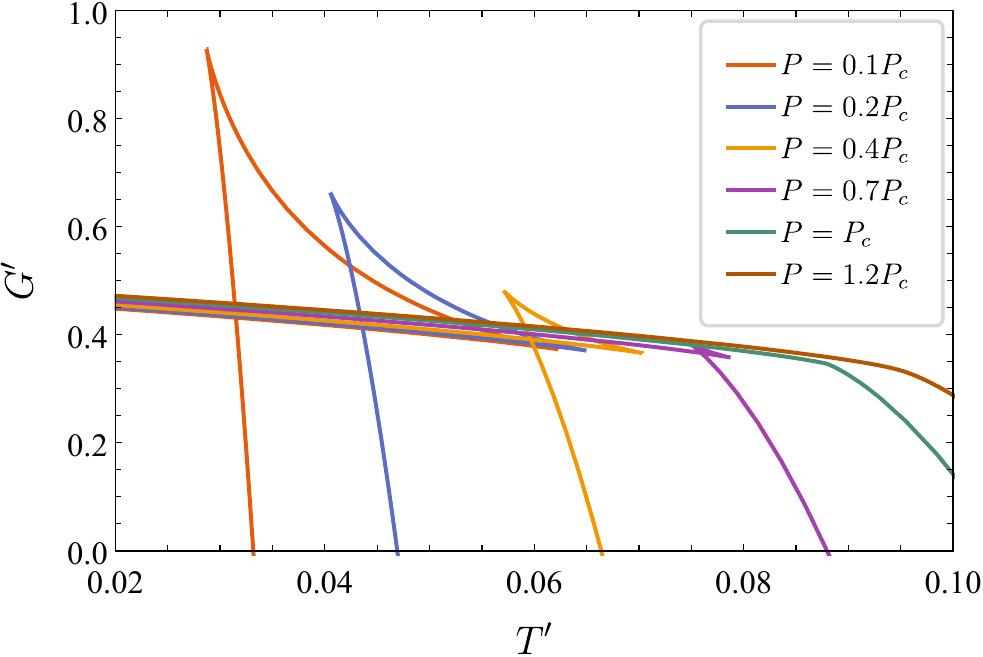}  
		\caption{}
		\label{fig:HayGT}
	\end{subfigure}
	\caption{(a) The coexistence curve of small-large phase and (b) the swallow tail behavior  on the $G' -T'$ plane of Hayward AdS black holes. }
	\label{fig:}
\end{figure}

In addition, the black dots are given by the new Gibbs free energy Eq.~\eqref{eq:G}, and the solid curve is given by the Maxwell equal area law in Fig.~\ref{fig:Hcoe}. The results computed by the two methods strictly  coincide with each other. The coexistence curve terminates at the critical point $(T'_c, P_c)$, where the area of a swallow tail equals  zero as shown in Fig.~\ref{fig:HayGP} and Fig.~\ref{fig:HayGT}.
As this critical point corresponds  to the point at which the first-order phase transition vanishes, we conclude that the Maxwell equal area law holds strictly for Hayward AdS black holes.

Similarly, in order to compare the Hayward AdS black hole with the van der Waals fluid,
we define the specific volume, $v\equiv 2r_+$, and rewrite the equation of state as follows:
\begin{eqnarray}
	P=\frac{T'}{v}-\frac{1}{2 \pi  v^2}+	\frac{8 q^3}{\pi  v^5}.
\end{eqnarray}
The critical condition Eq.~\eqref{Critical} determines the critical  values,
\begin{eqnarray}
	T'_c=\frac{3}{8 \sqrt[3]{20} \pi  q},\qquad v_c=	2 \sqrt[3]{20} q,\qquad P_c=\frac{3}{80 \sqrt[3]{50} \pi  q^2}.
\end{eqnarray}
Although these values are different from those in Eq.~(\ref{critvaltvp}), 
they give rise to the same ratio as Eq.~(\ref{gennum}).
As a result, this ratio is the same for Bardeen and Hayward AdS black holes, but is different from that of van der Waals fluids.

\subsection{A ratio of critical relations for some models of gravity theories}
\label{sec:number}

Inspired by the equal ratio in Bardeen and Hayward AdS black holes, we extend this issue to other models of gravity theories. In thermodynamics, each fluid corresponds to the critical relation, the critical pressure times the critical specific volume over the critical temperature, which is only related to the substance of which the fluid is composed. In black holes thermodynamics, we list the ratios of critical relations in Table~\ref{tab:number} by reviewing different models of gravity,  where we see that the ratios depend on the theories of gravity and the external fields coupling with gravity.

As mentioned above, the ratio of Bardeen AdS black holes equals that of Hayward AdS black holes because the two models belong to the same framework, i.e., the Einstein gravity coupled with nonlinear electrodynamics.  
We speculate that the ratio may keep unchanged for any other regular black holes in this framework. One reason comes from the ratio of charged AdS black holes  in the first two rows of Table~\ref{tab:number}.
Nonetheless, it is still a hypothesis that the ratio of $2/5$ is universal in Einstein's gravity coupled with nonlinear electrodynamics.

\renewcommand\arraystretch{1.5}
 \begin{table}[h]
	\centering
	\begin{tabular}{ p{4em} cc}
		\hline
		\hline
		 Number & Model   \\
		\hline
		${3}/{8}$ &	charged AdS black hole~\cite{Kubiznak:2012wp}  \\
		${3}/{8}$ &	charged AdS black hole with a global monopole~\cite{Deng:2018wrd}  \\
		${1}/{3}$ &	Gauss-Bonnet AdS black hole~\cite{Cai:2013qga}  \\
		$0.6742$ &	charged AdS black hole in conformal gravity~\cite{Xu:2014kwa}  \\
     	${5}/{12}$ &	black hole conformally coupled to scalar fields in $AdS_5$ spacetime~\cite{Miao:2016ieh}  \\
     	$({\lambda+1})/[{2(\lambda+2)}]$ &	AdS black hole in massive gravity~\cite{Fernando:2016sps}  \\
	    ${2}/{5}$ &	Bardeen AdS black hole  \\
	    ${2}/{5}$ &	Hayward AdS black hole  \\
		\hline
		\hline
	\end{tabular}
	\caption{The ratios of critical relations in some black hole models of gravity theories, where $\lambda$ is the constant in massive gravity.}\label{tab:number}
\end{table}

	

		
		
		
		
		
		
	

   \section{Summary} 
   \label{sec:con}
   
We revisit the thermodynamics of regular black holes in Einstein's gravity coupled with nonlinear electrodynamics, and establish an improved first law of thermodynamics by introducing a new extended phase space. In terms of this improved first law and new extended phase space, we recover the consistency of the Bekenstein-Hawking entropy formula and Maxwell equal area law for a general model in Einstein's gravity coupled with nonlinear electrodynamics. Then we apply the improved first law together with the new extended phase space to Bardeen and Hayward AdS black holes, and find that the inconsistency and deviation of thermodynamic quantities, such as entropy and volume, are eliminated. Our results meet the fact that Einstein's gravity coupled with nonlinear electrodynamics is still within the framework of general relativity, and thus the Bekenstein-Hawking entropy formula and Maxwell equal area law should be valid as they are in Einstein's gravity. 
Our work makes the misinterpretations clear, where such misinterpretations appeared in earlier literature on the thermodynamics of regular black holes in Einstein's gravity coupled with nonlinear electrodynamics.

In addition, we observe the equal ratio at the critical point of phase transitions for both Bardeen and Hayward AdS black holes. Our future task is to prove whether such a ratio is universal for any other regular black holes in Einstein's gravity coupled with nonlinear electrodynamics.

 \paragraph{Acknowledgments}
 The authors would like to thank  the anonymous referee for the helpful comments that improve this work greatly.
 This work was supported in part by the National Natural Science Foundation of China under Grant No. 12175108. 
      

\bibliographystyle{JHEP}
\bibliography{references}

\end{document}